\documentclass[aps,pre,onecolumn,preprintnumbers,superscriptaddress,longbibliography,10pt]{revtex4-2}
\usepackage{import}
\usepackage[T1]{fontenc}
\usepackage[framemethod=tikz]{mdframed}
\usepackage[english]{babel}
\usepackage[utf8]{inputenc}
\usepackage{fancyhdr}
\usepackage{float}
\usepackage{graphicx}
\usepackage[us,12hr]{datetime}
\usepackage{marvosym}
\usepackage{booktabs}
\usepackage{amsmath,amssymb,amsbsy,amstext,amsthm,simplewick}
\usepackage{graphicx,float}
\usepackage{color}
\usepackage{xcolor}
\usepackage{mathrsfs}
\usepackage{aas_macros,empheq,fancybox,multirow}
\usepackage{slashed}
\usepackage{braket}
\usepackage{dsfont}
\usepackage{nccmath}
\usepackage{scalerel}
\usepackage{url}
\usepackage{tikz}
\usepackage[unicode=true,
            pdfusetitle,
            colorlinks=true
            ,urlcolor=mn
            ,anchorcolor=black
            ,citecolor=magenta
            ,filecolor=navyblue
            ,linkcolor=rindou2
            ,menucolor=navyblue
            ,linktocpage=true
            ,pdfproducer=medialab]{hyperref}
\hypersetup{pdfauthor={Name}}

\definecolor{rindou1}{rgb}{0.4431,0.2862,0.7960}
\definecolor{rindou2}{rgb}{0.0078,0.1215,0.4392}
\definecolor{lapis}{rgb}{0.0.0470,0.2941,0.5568}
\definecolor{mn}{rgb}{0.15, 0.35, 0.95}
\definecolor{br}{HTML}{792500}
\definecolor{br}{HTML}{C70039}

\definecolor{lime}{HTML}{A6CE39}
\DeclareRobustCommand{\orcidicon}{%
	\begin{tikzpicture}
	\draw[lime, fill=lime] (0,0) 
	circle [radius=0.16] 
	node[white] {{\fontfamily{qag}\selectfont \tiny ID}};	\draw[white, fill=white] (-0.0625,0.095) 
	circle [radius=0.007];	\end{tikzpicture}
	\hspace{-2mm}}
\foreach \x in {A,...,Z}{%
	\expandafter\xdef\csname orcid\x\endcsname{\noexpand\href{https://orcid.org/\csname orcidauthor\x\endcsname}{\noexpand\orcidicon}}
}

\begin{document}
\title{Phase diagram of generalized XY model using tensor renormalization group}

\author{Abhishek Samlodia\orcidC{}}
\affiliation{Department of Physics, Syracuse University, Syracuse NY 13244, USA}

\author{Vamika Longia\orcidD{}}
\affiliation{Department of Physical Sciences, Indian Institute of Science Education Research - Mohali, Knowledge City, Sector 81, SAS Nagar, Punjab 140306, India}

\author{Raghav G.~Jha\orcidA{}}
\affiliation{Thomas Jefferson National Accelerator Facility, Newport News, VA 23606, USA}

\author{Anosh Joseph\orcidB{}}
\affiliation{National Institute for Theoretical and Computational Sciences, School of Physics, and Mandelstam Institute for Theoretical Physics, University of the Witwatersrand, Johannesburg, Wits 2050, South Africa}

\begin{abstract}
We use the higher-order tensor renormalization group method to study the two-dimensional generalized XY model that admits integer and half-integer vortices. This model is the deformation of the classical XY model and has a rich phase structure consisting of nematic, ferromagnetic, and disordered phases and three transition lines belonging to the Berezinskii-Kosterlitz-Thouless (BKT) and Ising class. We explore the model for a wide range of temperatures, $T$, and the deformation parameter, $\Delta$, and compute specific heat along with integer and half-integer magnetic susceptibility, finding both BKT-like and Ising-like transitions and the region where they meet. 
\end{abstract}

\maketitle


\section{Introduction}
\label{sec:intro}

Spin models in two Euclidean dimensions with discrete or continuous global symmetries can have a wide range of interesting properties. Due to the famous no-go theorem known as the Hohenberg-Mermin-Wagner-Coleman (HMWC) theorem, in two-dimensional models, a continuous symmetry cannot break spontaneously, and thus, a phase transition from a disordered to an ordered phase is not allowed. The simplest model in two dimensions with continuous symmetry is the classical XY model, also known as the O(2) model. In this model, the spins 
take values on the circle $S^1$. This model has been the subject of several investigations over the past fifty years \cite{Berezinskii:1970fr, Kosterlitz:1973xp, Tobochnik79, S_Ota_1992, Mattis1984} due to its simplicity and remarkable properties. Surprisingly, it admits an infinite-order phase transition, which is topological, known as the Berezinskii-Kosterlitz-Thouless (BKT) phase transition. This transition is peculiar since it does not follow the usual classification of the phase transitions due to Ehrenfest, which describes the order of the phase transition in terms of the lowest discontinuous derivative of the partition function. 
This transition does not violate the HMWC theorem since the transition is not due to breaking of any symmetry but due to the unbinding of vortices and antivortices (topological defects) at some finite temperature. Across the phase transition in the XY model, all the derivatives of the free energy remain continuous. The transition is associated with the dissociation of the integer vortex pairs at the critical temperature. Due to the wide-ranging applications of this model, in explaining various phenomena related to superfluid Helium, thin films, superconductivity, liquid crystals, and melting of two-dimensional crystals, several extensions of this model have been considered. 

\begin{figure}
    \centering
    \includegraphics[width=8.5cm]{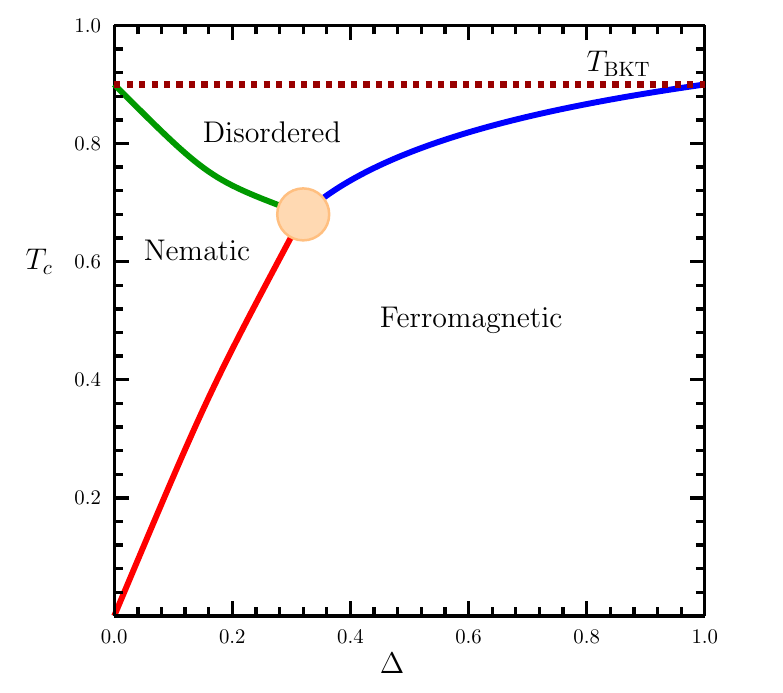}
    \caption{The conjectured phase diagram of the gXY model. The BKT line (blue) separates the ferromagnetic and the disordered phases, the half-BKT line (green) separates the nematic and disordered phases, whereas the Ising line (red) separates the nematic and ferromagnetic phases. The blob represents the region where the transition lines meet.}
    \label{fig:cartoon}
\end{figure}

One such extension we study here was first proposed by Korshunov, Lee, and Grinstein (KLG) in Refs.~\cite{Korshunov85, Lee85}. This model has several interesting features, some of which include the possibility of fractional vortices and signs of passing directly from the disordered (high temperature) phase to the single particle (quasi) condensate phase via an Ising transition, a situation reminiscent of the `deconfined criticality' scenario as studied in Ref.~\cite{Shi2011}. Due to the competition between the different terms in the Hamiltonian, this also leads to a richer phase structure. This model also provides a good example to understand the interplay between the discrete $\mathbb{Z}_{2}$ symmetry and $U(1)$ symmetry. The modification of the XY model, as proposed by KLG, is given by the Hamiltonian:
\begin{equation}
\label{eq:gXY_ham} 
\mathcal{H} = -J \sum_{\langle jk \rangle} \cos(\theta_j - \theta_k) - J_1  \sum_{\langle jk \rangle} \cos(q(\theta_j - \theta_k)), 
\end{equation}
where we use the standard notation $\langle jk \rangle$ to denote the nearest-neighbor and $\theta_j \in [0, 2\pi)$ with $J, J_1 > 0$. We can consider these couplings to depend on a parameter $\Delta$ such that $J = \Delta$ and $J_1 = 1 - \Delta$ with $0 \le \Delta \le 1$. The limit $\Delta = 0$ corresponds to a pure spin-nematic model, while $\Delta = 1$ is the usual XY model. The presence of the nematic term $\cos(q\theta_j - q\theta_k)$ gives rise to fractional excitations such as half-integer vortices (for $q = 2$), and these exhibit invariance under $\theta \to \theta + (2\pi /q)$. In this work, we will only consider the case $q = 2$. The model defined by \eqref{eq:gXY_ham} is known as the generalized XY (gXY) model, and the phase diagram of this model has been subject to a lot of investigations. The basic conclusion is that there are three phases denoted by BKT, half-BKT, and Ising-like transitions \cite{DB_Carpenter_1989, Lee85, Hubscher2012, Canova2016, Serna_2017} and they connect~\cite{Nui2018} around a special point in the $T-\Delta$ phase space. We show the expected phase structure in Fig.~\ref{fig:cartoon}. In this work, we perform numerical computations to precisely sketch this diagram, using the numerical tensor network method, finding evidence that the Ising transition line might preempt the BKT transition for a finite range of $\Delta$. 

When $\Delta = 0$, there is a topological transition corresponding to the dissociation (unbinding) of half-integer vortex and anti-vortex, while for $\Delta = 1$, there is a dissociation of integer vortex-pairs. Both these transitions happen around the same temperature~\cite{DB_Carpenter_1989}. Though the qualitative behavior of the phase diagram of gXY model is known, a determination of the multi-critical point or how the transition lines connect is mostly known due to Monte Carlo numerical work. The goal of this work is to study the phase diagram by exploring several $\Delta$ values for the case of $q = 2$ and provide an alternate method using real-space tensor renormalization group techniques by approximating the partition function corresponding to the gXY Hamiltonian. There has also been a lot of work for $q > 2$ finding interesting additional phases. We refer the reader to Ref.~\cite{Poderoso2011} to start the reference trail. 

The study of phase transitions and symmetry breaking for the case of statistical models with short-range interactions in two dimensions is special due to the famous HMWC theorem, which states that continuous symmetry cannot break spontaneously. An equivalent statement is that there are no Goldstone bosons that accompany the symmetry breaking in two Euclidean dimensions. Due to this, the XY model is interesting in itself, but adding the additional term makes the model even more interesting. The richness of this model can be understood as follows: there are three primary phase regions - a disordered region, a region with integer vortices (IV), and a region with half-integer vortices (HIV). The integer-vortex pair phase is also referred to as the `ferromagnetic,' whereas the `nematic' represents the half-integer vortex pair phase. The transition line, corresponding to a continuous phase transition, also referred to as the `Ising-line,' can be captured by the logarithmic divergence of the specific heat close to the critical temperature, while it is known that such a method is not very useful for BKT-like transitions. To locate the transition between the integer-vortex pair phase to the disordered phase (where the vortices dissociate), we compute the magnetization for a small symmetry-breaking external field and then take the vanishing field limit to extract the critical temperature. We cannot compute the spontaneous magnetization in the thermodynamic limit without an external field since, due to the HMWC theorem, it is zero. This method of determining the infinite-order transition was pursued in Ref.~\cite{Jha:2020oik}. For the IV to HIV, the transition can simply be tracked by locating the divergence in specific heat since it is of Ising-type.
 
To carry out a precise study of the phase transitions in this model, we apply the tensor network methods based on the higher-order tensor renormalization group (HOTRG) as introduced in \cite{Xie2012}. Some preliminary investigation of this model was carried out in \cite{Longia:2023dus}. This model has been studied using the variational uniform matrix product state (VUMPS) algorithm \cite{Zauner-Stauber:2017eqw} in Ref.~\cite{Song_2021}. Here, we use an alternative approach using the numerical real-space tensor renormalization group methods. Since the proposal by Levin and Nave \cite{Levin:2006jai}, the scheme of performing real-space coarse-graining using tensors has seen a lot of progress with extension to higher dimensions and innovative procedures to carry out the tensor renormalization ~\cite{Kadoh:2019kqk, Adachi2020, Az-zahra:2024gqr}. We refer the interested reader to Ref.~\cite{Ran2020} for summary of tensors networks based on both real-space renormalization group (RG) methods and approximation of the ground state of slightly entangled many-body systems. Tensor renormalization group (TRG) approach has been successfully applied to the two-dimensional XY model and related models \cite{Yu:2013sbi, Akiyama:2019xzy, Jha:2020oik, Hostetler:2021uml, Jha:2022pgy, Butt:2022qqx} and also to three-dimensional XY model with finite chemical potential \cite{Bloch:2021mjw}. Using the dual variable approach (from the character expansion), several gauge theories have also been studied \cite{Bazavov:2019qih, Akiyama:2022eip, Kuwahara:2022ubg} with tensors and have found agreement with results obtained using other numerical methods. In addition, recently some progress was also made to extract the critical behavior by studying the renormalization group collapse of magnetization for three-dimensional $O(4)$ model \cite{Akiyama:2024qgv}. It seems likely that once can compute critical exponents accurately using TRG methods in coming years providing an alternative to the existing methods such as Monte Carlo and conformal bootstrap. The major part of these computations is the contraction of the network during coarse-graining and the 
singular value decomposition (SVD) of the growing size of tensors to restrict to fixed size after each step. This amounts to truncating the singular values to some threshold $D$ which is known as bond dimension. To improve the performance of the numerical computations in this work and get to $D=91$ in a reasonable amount of computational time, we use the grpahical processing unit (GPU)-accelerated code described in Ref.~\cite{Jha:2023bpn}.

\section{Tensor formulation}
\label{sec:TF}

The starting point of the real-space tensor renormalization approach (on square lattice) is to decompose the Boltzmann weight in terms of tensor with $2d$ indices where $d$ is the number of dimensions (Euclidean). This can either be done exactly or approximately based on the symmetries of the action. One of the standard methods to do this decomposition is to use the character expansion \cite{Liu:2013nsa}, but different truncation schemes also exist \cite{Kadoh:2018tis}. Using this initial tensor one can create a network of these tensors such that when the network is contracted, it provides a good approximation to the partition function $Z$. 

The partition function of the classical spin model with external field $h$ is:  
\begin{equation}
\begin{split}
    Z = \prod_j \int \frac{d\theta_j}{2\pi} \prod_{\langle jk \rangle} e^{\beta [\Delta \cos(\theta_j - \theta_k) + (1 - \Delta) \cos(2(\theta_j - \theta_k))]} \times \prod_j e^{\beta [h \cos(\theta_j)]},
\end{split}
\end{equation}
where $\langle jk \rangle$ denotes neighboring lattice sites and $\beta$ is the inverse temperature. We expand the argument of the exponential (Euclidean action) using Jacobi-Anger expansion and obtain: 
\begin{equation}
\begin{split} 
    Z = \prod_j \int \frac{d\theta_j}{2\pi} \prod_{l\in L} \sum_{n_l} a_{n_l}(\beta, \Delta)e^{i n_l(\theta_{j} - \theta_{k})} \times \sum_{p_l} I_{p_{l}}(\beta h) e^{i p_l \theta_{j}}, 
\end{split}
\end{equation}
where
\begin{equation}
\label{eq:a_n} 
    a_n (\beta, \Delta)= \sum_{m = -\infty}^\infty I_{n - 2m}(\beta \Delta) I_m(\beta(1 - \Delta)),
\end{equation}
and $I_n$ is the modified Bessel function of the first kind. On integrating over the $\theta_j$ variables, we obtain the partition function:  
\begin{equation}
\label{eq:iniT} 
    Z \approx {\rm tTr}\left( \prod_s T_{n_1, n_2, n_3, n_4}(s) \right),
\end{equation}
where
\begin{equation}
\label{eq:Tijkl} 
\begin{split}
    T_{n_1, n_2, n_3, n_4}(s) = \sqrt{\prod_{k = 1}^4a_{n_k}(\beta, \Delta)} \times I_{n_1 + n_2 - n_3 - n_4}(\beta h).
\end{split} 
\end{equation}
This tensor represents the four-legged object at site $s$ where $n_1$, $n_2$, $n_3$, and $n_4$ denote the top, right, bottom, and left legs, respectively. In the case of zero field, i.e., $h = 0$, we have the conservation of $U(1)$ charges due to the $\delta$-function in \eqref{eq:Tijkl}. To write the initial tensor for numerical computations, we have to choose a suitable range of values for $m$ so that the infinite sum in ~\eqref{eq:a_n} can be truncated to a finite interval. We have fixed this to be integral values, $m \in [-50, 50]$. The second truncation we have to do is over the indices in \eqref{eq:Tijkl}, which also runs from $-\infty$ to $\infty$. For this truncation, we choose $n_k \in [-40, 40], \; n_k \in \mathbb{Z}$, i.e., $D = 91$. Using these truncation procedures, we use the initial tensor to perform a fixed number of coarse-graining steps, $N = 30$ for most cases, using the higher-order singular value decomposition of the tensors (HOTRG algorithm) as described in ~\cite{Yu:2013sbi, Jha:2020oik} to obtain the partition function in the thermodynamic limit. We refer the reader to  Appendix~\ref{app:AppA} for additional details about the tensor formulation of the generalized XY model and the impure tensor for magnetization. 

\section{Numerical  Results}
\label{sec:results}

We now discuss the results for the generalized XY model using tensor network methods. The main quantity of interest in the real-space tensor computation is the partition function, which has to be approximated accurately. Using this, we compute two main observables - the specific heat and the magnetic susceptibility normalized by the lattice volume $V$. They are defined as: 
\begin{equation}
\label{eq:spech}
     C_{\rm{v}} = \frac{\beta^2}{V} \frac{\partial^2{\ln Z}}{\partial \beta^2},
\end{equation}
and
\begin{equation}
\label{eq:susc}
    \chi = \frac{1}{V}\frac{\partial M}{\partial h} \; = \; \frac{1}{\beta V}\frac{\partial^2 \ln Z}{\partial h^2},
\end{equation}
where $M$ is the magnetization, which is defined using the free energy, $F = -T \ln(Z) = -\beta^{-1} \ln(Z)$ as:
\begin{equation}\label{eq:mag}
    M = -\frac{\partial F}{\partial h} = \frac{1}{\beta} \frac{\partial {\rm ln} Z}{\partial h}. 
\end{equation}
The computation of magnetization from the first derivative of free energy (or partition function) is often prone to numerical errors. A better method is to compute the impure tensor corresponding to the magnetization and insert it into the network for the partition function~\cite{Jha:2020oik}. To compute the dominant magnetization, depending on $\Delta$, we use either $h$ or $h_{1}$ corresponding to the different symmetry-breaking terms in the partition function. We use external magnetic field $h$ for $\Delta > 0.36$  to compute $M$ and magnetic field $h_{1}$ for $\Delta \le 0.36$ related to $\cos{(2\theta)}$, term which we refer to as $M_{1}$. We refer the reader to Appendix~\ref{app:AppA} for further details about the tensor construction. 

Since this model has Ising and BKT-like transitions, the specific heat cannot always conclusively determine the phase transition. For such cases, we look at the first derivative of the magnetization computed using a simple finite difference method and compute the susceptibility. The peak in susceptibility signals a transition, which we then extrapolate to the zero-field limit. The extraction of this zero field critical temperature is obtained by doing functional fits of the form discussed in Ref.~\cite{Jha:2020oik}. We show the magnetic susceptibility plot and zero-field limit extraction of the critical temperature for $\Delta = 0.32$ in Fig.~\ref{fig:chi0p32} and Fig.~\ref{fig:Tc0p32}, respectively, where $\overline{h}_1$ is the central value used for the finite-difference of the susceptibility measurement. We refer the reader to Appendix \ref{app:AppC} for plots corresponding to a wide range of $\Delta$ values and to Appendix \ref{app:AppD} for plots corresponding to systematic error analysis when $m$ or $D$ is varied. For the computation of magnetization, we use a lattice volume of $2^{35} \times 2^{35}$ while for $h = h_{1} = 0$, we use a volume of $2^{30} \times 2^{30}$. For all computations, we use $D = 91$ and a range of $m$ as defined in \eqref{eq:a_n} to be $[-50,50]$.

\begin{minipage}{0.9\linewidth}
      \centering
      \begin{minipage}{0.45\linewidth}
          \begin{figure}[H]
              \includegraphics[width=\linewidth]{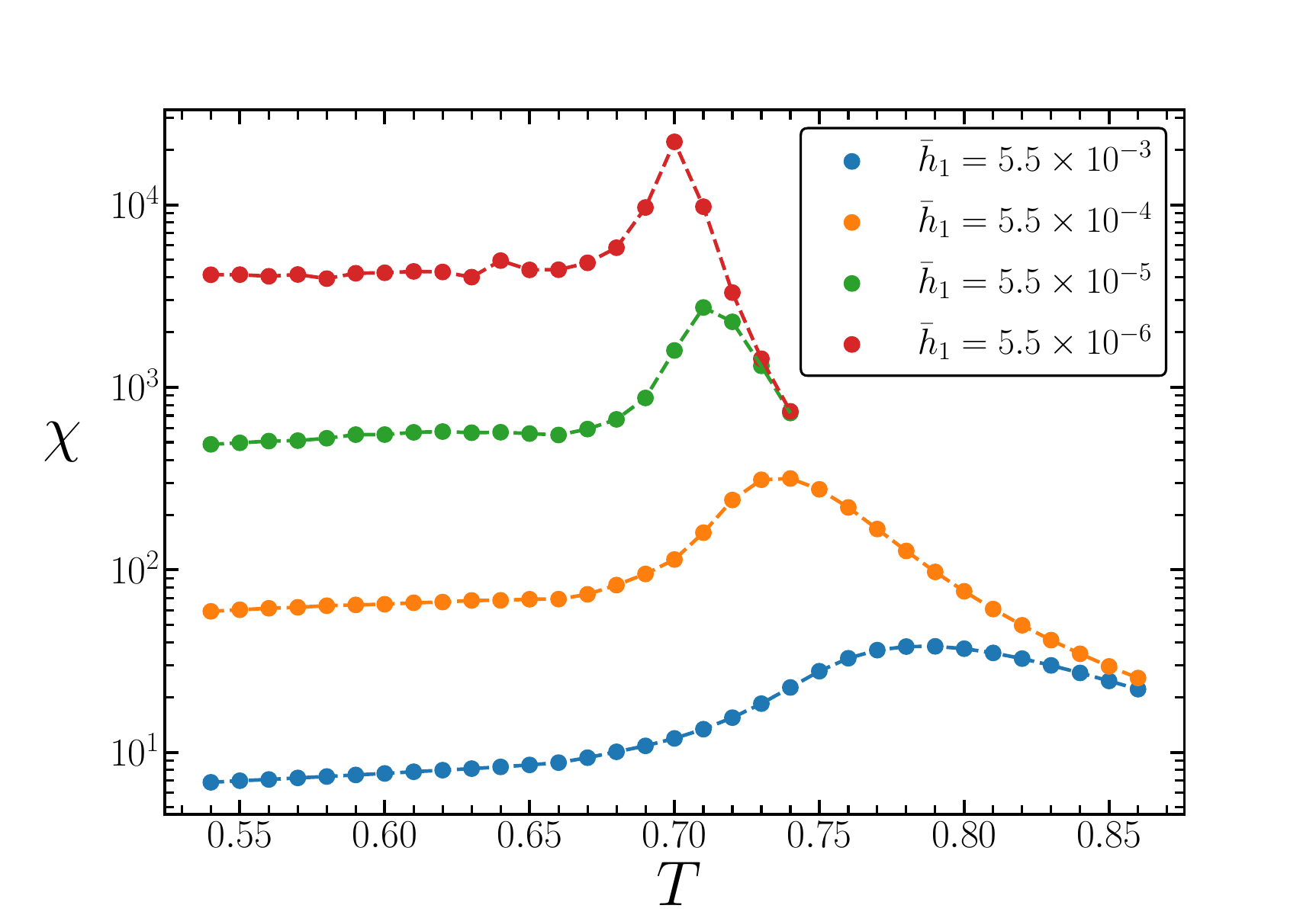}
              \caption{\label{fig:chi0p32}The magnetic susceptibility defined in \eqref{eq:susc} against $T$ for $\Delta = 0.32$.}
          \end{figure}
      \end{minipage}
      \hspace{0.05\linewidth}
      \begin{minipage}{0.44\linewidth}
          \begin{figure}[H]
              \includegraphics[width=\linewidth]{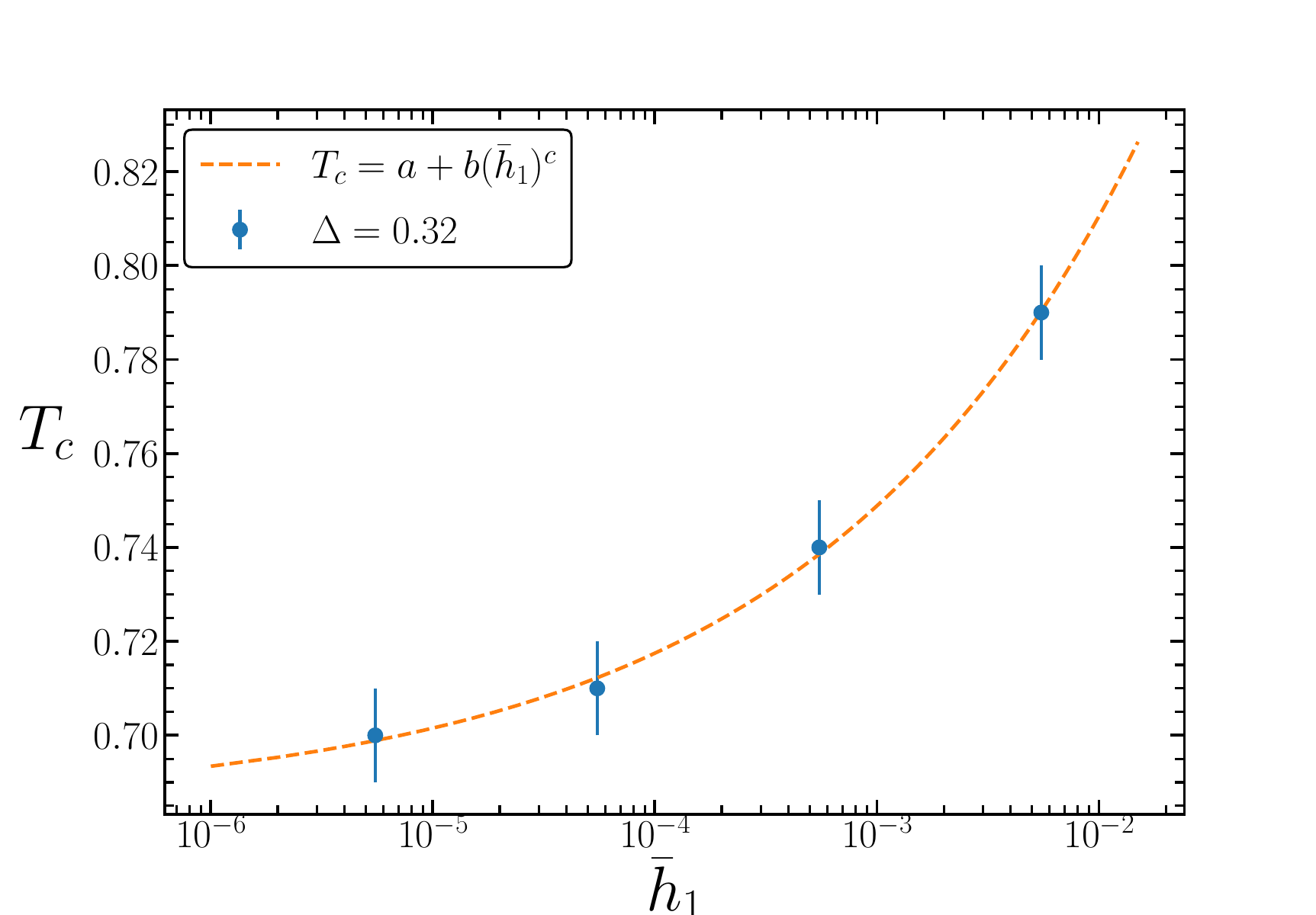}
              \caption{\label{fig:Tc0p32}The critical temperature, $T_c$ for \\
              different $\overline{h}_{1}$ and $\Delta = 0.32$. $a = 0.685(3), b = 0.48(1), c = 0.29(5)$.}
          \end{figure}
      \end{minipage}
  \end{minipage}
\\
\\
\\
As discussed before, the gXY model reduces to the XY model when $\Delta = 1$, and much work has been done using Monte Carlo and tensor methods for the computation of the critical temperature and critical exponents. The critical temperature on a square lattice was computed to be $T_c = 0.89290(5)$ using tensor network methods in \cite{Jha:2020oik}, while the critical exponent $\delta \approx 15$ was computed within errors in Ref.~\cite{Jha:2023bpn}. In this work, we start with $\Delta = 0.8$, which is expected to have a single BKT-like transition, and move to smaller $\Delta$ values passing through the region where the transition lines meet and going all the way down to the limit of $\Delta \to 0$. For the BKT transition, it is well-known that the peak of specific heat is not the correct way to determine the transition temperature (it is an overestimate, as clear from the data in Table~\ref{table1}). From our computation, we find that for $\Delta = 0.8$, the peak of $C_{\rm{v}}$ is observed at $T = 0.95(1)$. This was reported to be around $T \approx 0.91$ in Ref.~\cite{Song_2021}. To accurately determine the transition, we compute magnetic susceptibility for a small external field $h$ and then took the zero-field limit as explained above. For $\Delta = 0.8$, we obtain $T \sim 0.890(4)$. 

As we decrease $\Delta$ approaching the multicritical point (or the region where the phases meet), the difference between the transition temperatures deduced from the peak of specific heat and magnetic susceptibility, respectively, decreases for the BKT transition. As we cross the region where transition lines meet and move to smaller values of $\Delta$, we again see that the difference increases for the half-BKT line. This is evident from the data given in Table~\ref{table1} in Appendix~\ref{app:AppB}. Such behavior has also been noted previously in Ref.~\cite{Hubscher2012} where for the half-BKT line, the critical temperature values almost agree with each other around $\Delta = 0.35$. We find that the specific heat peak fails to capture the correct transition temperature for $\Delta \ge 0.50$. 

\begin{figure}
    \centering
    \includegraphics[scale=0.45]{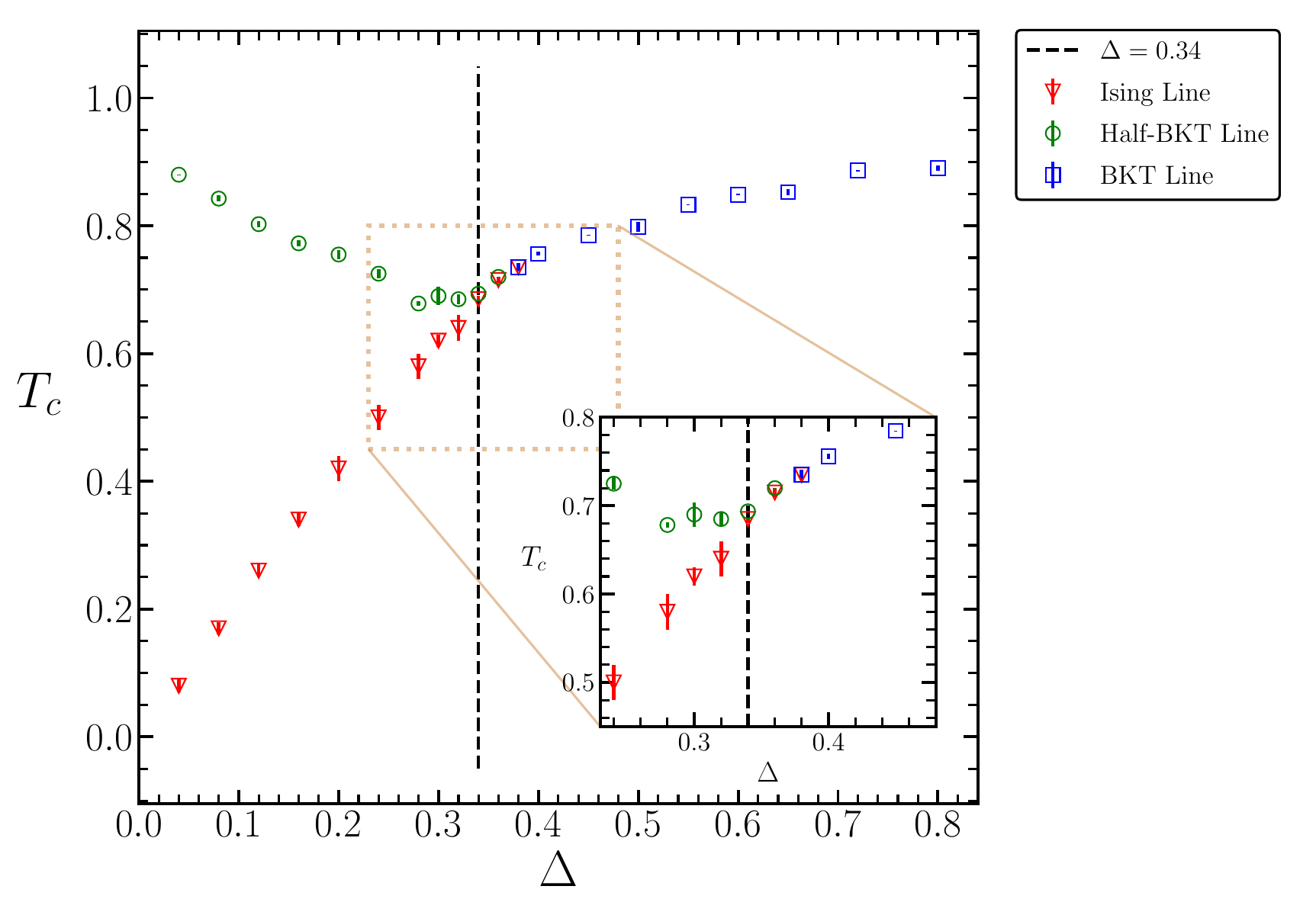}
    \caption{The critical temperatures for a range of deformation parameter $\Delta$ obtained using TRG. The red triangles represent the continuous Ising transition, while the green circles and blue squares represent the transition of the BKT class.}
    \label{fig:phase_diag}
\end{figure}

From the results of our tensor network computations, we find the phase diagram for this model as shown in Fig.~\ref{fig:phase_diag}. The critical temperatures for the half-BKT and BKT lines are extracted from the zero-field limit method, as stated before, and the critical temperatures for the Ising line are inferred using the peak in the specific heat. We collect the numerical results in Table~\ref{table1} for critical temperature values deduced from the peak of the specific heat as well as magnetic susceptibility corresponding to the standard and nematic magnetization. In Appendix~\ref{app:AppC}, Fig.~\ref{fig:chi_T1}-\ref{fig:delta_T5} show the magnetic susceptibility v/s temperature plots as well as critical temperature scaling with finite external magnetic field for a variety of $\Delta$ values. In Appendix~\ref{app:AppD}, Table~\ref{table2} we show the variation of free energy and magnetization $M_1$ as one varies $m$ parameter at fixed $\Delta = 0.34$ near the critical temperature and Fig.~\ref{fig:scale1}-\ref{fig:scale3} show the free energy scaling with increasing bond-dimension near the transition point different deformation parameter values.

\section{Conclusions} 
\label{sec:Conclusions}

We studied the phase diagram of the generalized XY model in two dimensions using the GPU-improved real-space tensor network methods. We located the phase transitions belonging to the Ising and the BKT universality classes and determined the region in the phase diagram where they appear to connect, i.e., $\Delta = 0.36(2), T_c = 0.716(3)$. Our result is consistent with an interpretation where the half-BKT and Ising transition lines first meet (as $\Delta$ is increased from zero), and then the Ising line continues for a small range of $\Delta$ to merge with the BKT line, as observed in Ref.~\cite{Serna_2017} for $q=2$ and for $q=3$ Potts model in Ref.~\cite{Drouin-Touchette2022}. This work refines the previous computation of this model using matrix product states (MPS) methods, and we find results that are qualitatively similar to the existing results in the literature from the standard Monte Carlo method. It remains an open problem to apply and extend the methods used in this work to models involving fractional vortices, corresponding to terms like $\cos(p\theta_i - p\theta_j)$ with $p \ge 3$ in the Hamiltonian. It is believed that these models have a more complicated phase diagram, as studied in Ref.~\cite{Poderoso2011} using Monte Carlo methods. It would be interesting to study and revisit this model with the tensor network methods in the future. 

\subsection*{Acknowledgments}
A.S. is supported by U.S. Department of Energy grant DE-SC0019139. R.G.J. is supported by the U.S. Department of Energy, Office of Science, National Quantum Information Science Research Centers, Co-design Center for Quantum Advantage (C2QA) under contract number DE-SC0012704 and by the U.S. Department of Energy, Office of Science, Office of Nuclear Physics under contract number DE-AC05- 06OR23177.
The work of A.J. was supported in part by the Start-up Research Grant from the University of the Witwatersrand. The numerical computations were done on PARAM~SMRITI, Mohali, India, supported by the Ministry of Electronics and Information Technology and Department of Science and Technology (DST), Government of India, and Syracuse University HTC Campus Grid supported by NSF award ACI-1341006. We thank Minati Biswal for collaboration during the initial stages of this work. 

\subsection*{Data Availability Statement}
The code and data used in this paper can be obtained from Ref.~\cite{Samlodia2024}.

\appendix

\section{\label{app:AppA}Derivation of tensors for the generalized O(2) model}

We start with the following Hamiltonian
\begin{equation}
    \begin{split}
        H = & -\Delta \sum_{\langle j k \rangle} \cos{(\theta_j - \theta_k)} - (1-\Delta) \sum_{\langle j k \rangle} \cos{(2(\theta_j - \theta_k))}
         - h \sum_j \cos{(\theta_j)} - h_1\sum_j \cos{(2\theta_j)},
         \label{eq:appendixH} 
    \end{split}
\end{equation}
which represents the generalized XY model with the symmetry-breaking fields corresponding to the integer and half-integer terms. The partition function then reads
\begin{equation}
    \begin{split}
        Z = & \int \Bigg[\prod_j \frac{d\theta_j}{2\pi}\Bigg] \Bigg(e^{\beta\Delta \sum_{\langle j k \rangle} \cos{(\theta_j - \theta_k)} + \beta h\sum_j \cos{(\theta_j)}} \times
        e^{\beta(1 - \Delta) \sum_{\langle j k \rangle} \cos{(2(\theta_j - \theta_k))} + \beta h_1 \sum_j \cos{(2\theta_j)}}\Bigg).
    \end{split}
\end{equation}
Using the expansion in terms of the dual variables
\begin{equation}
    e^{\beta\Delta\cos{(\theta_j - \theta_k)}} = \sum_{a = -\infty}^\infty I_a(\beta\Delta)e^{ia(\theta_j - \theta_k)},
\end{equation}
where $I_a$ is the modified Bessel function of the first kind with integral order $a$. If we write a summation for each term in the Hamiltonian and then collect all the terms corresponding to $\theta_j$, we can write the partition function as a tensor network contraction of four-ranked tensors living on each lattice site. Such a tensor at lattice site $s$ is given by
\begin{equation}
\begin{split}
    T^s_{abcd} = &\sqrt{ \sum_{m,n,o,p} I_{a}(\beta\Delta)I_{b}(\beta\Delta)I_{c}(\beta\Delta)I_{d}(\beta\Delta)}\; \times \sqrt{I_m(\beta(1-\Delta))I_n(\beta(1-\Delta))I_o(\beta(1-\Delta))I_p(\beta(1-\Delta))} \; \times \\ & \sum_{r} I_r(\beta h_1)\sum_{l}I_l(\beta h) \int \frac{d\theta_s}{2\pi} \;\times e^{i\theta_s((a - b + c - d) + 2(m - n + o - p) + l + 2r)}. \\
\end{split}
\end{equation}
To evaluate this integral, we can relabel the indices as follows:
\begin{equation}
    \begin{split}
        a ~\to~ & a - 2m, \\
        b ~\to~ & b - 2n, \\
        c ~\to~ & c - 2o, \\
        d ~\to~ & d - 2p. \\
    \end{split}
\end{equation}
Such a relabeling does not affect the integral since the indices of the Bessel functions run from $-\infty$ to $+\infty$. The tensor $T^s_{abcd}$ becomes: 
\begin{equation}
\begin{split}
    T^s_{abcd} = &\sqrt{ \sum_{m,n,o,p} I_{a-2m}(\beta\Delta)I_{b-2n}(\beta\Delta)I_{c-2o}(\beta\Delta)I_{d-2p}(\beta\Delta)}\; \times \sqrt{I_m(\beta(1-\Delta))I_n(\beta(1-\Delta)) I_o(\beta(1-\Delta))} \; \times\\ 
    & \sqrt{I_p(\beta(1-\Delta))} \; \sum_{l} I_l(\beta h) \sum_{r}I_r(\beta h_1) \int \frac{d\theta_s}{2\pi} e^{i\theta_s((a - b + c - d) + l + 2r)}. \\
\end{split}
\end{equation}
The integral is the familiar Fourier transform of the delta function; hence contracting all the other terms with this delta function, we get the final form of the four-ranked site tensor as
\begin{equation}
\begin{split}
    T^s_{abcd} = &\sqrt{ \sum_{m,n,o,p} I_{a-2m}(\beta\Delta)I_{b-2n}(\beta\Delta)I_{c-2o}(\beta\Delta)I_{d-2p}(\beta\Delta)}\; \times \sqrt{I_m(\beta(1-\Delta))I_n(\beta(1-\Delta)) I_o(\beta(1-\Delta))}\; \times \\
    & \sqrt{I_p(\beta(1-\Delta))} \; \sum_{r} I_{a - b + c - d + 2r}(\beta h) I_r(\beta h_1). \\
\end{split}
\end{equation}
To write the site tensor in a more compact notation, we define
\begin{equation}
    a_{n_k}(\beta, \Delta) = \sum_{\nu_k = -\infty}^\infty I_{n_k - 2\nu_k}(\beta\Delta) I_{\nu_k}(\beta(1-\Delta)),
\end{equation}
using which the tensor becomes:
\begin{equation}
    \begin{split}
        T_{n_1, n_2, n_3, n_4} = & \sqrt{\prod_{k = 1}^4 a_{n_k}(\beta, \Delta)}\; \times \sum_{l = -\infty}^\infty I_{n_1 + n_2 - n_3 - n_4 + 2l}(\beta h) I_l(\beta h_1).
        \label{eq:App_T_tensor}
    \end{split}
\end{equation}
The partition function can be approximated using this tensor description as a trace of the network:
\begin{equation}
    Z \approx {\rm tTr} \Big(\prod_s T_{n_1, n_2, n_3, n_4}(s)\Big).
    \label{eq:Z}
\end{equation}
Here, ${\rm tTr}$ implies a tensor trace of the tensor network. We show the site tensor and the partition function as a fully contracted tensor network consisting of $4 \times 4$ lattice in Fig. \ref{fig:T_abcd} and Fig. \ref{fig:tensor_network}, respectively.

\begin{minipage}{0.9\linewidth}
      \centering
      \begin{minipage}{0.4\linewidth}
          \begin{figure}[H]
              \includegraphics[width=\linewidth]{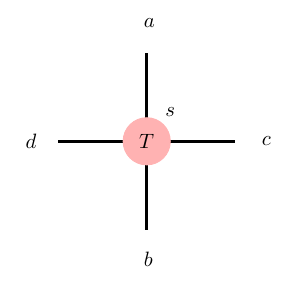}
              \caption{The rank four site tensor. We set $a(n_1)$ and $c(n_2)$
              indices positive and $b(n_3)$ and \\$d(n_4)$ indices negative as mentioned in 
              \eqref{eq:App_T_tensor}}
              \label{fig:T_abcd}
          \end{figure}
      \end{minipage}
      \hspace{0.05\linewidth}
      \begin{minipage}{0.37\linewidth}
          \begin{figure}[H]
              \includegraphics[width=\linewidth]{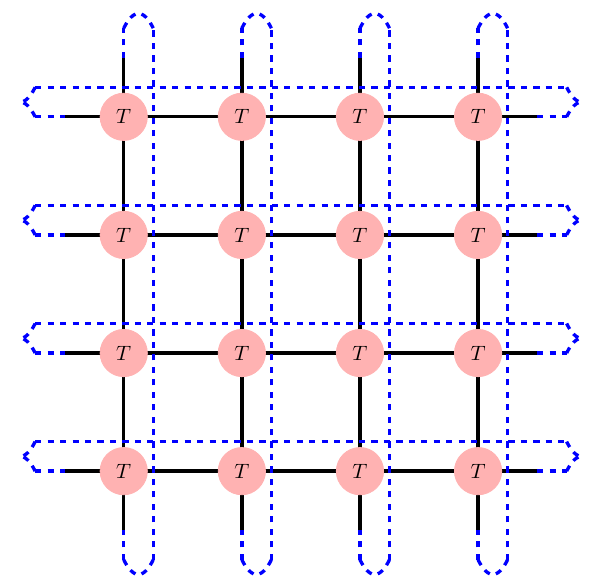}
              \caption{A fully contracted tensor network $T$ with periodic boundary condition that gives the partition function $Z$ on $4 \times 4$ lattice. Note that all the $T$ are the same due to translational symmetry.}
              \label{fig:tensor_network}
          \end{figure}
      \end{minipage}
\end{minipage}
\\
\\
\\
The magnetization is calculated using \eqref{eq:mag} and is $M = P/Z$, where $Z$ is the partition function given in \eqref{eq:Z} and $P$ is the modified contracted tensor network with impure tensor inserted. The impure tensor corresponding to $M$ is given by:
\begin{equation}
    \begin{split}
        \mathcal{I}_{n_1 n_2 n_3 n_4}^{M} = &\;  \left[\sqrt{\prod_{k = 1}^{4} a_{n_k}(\beta,\Delta)} \;\times\; \sum_{l = -\infty}^{+\infty} I_l(\beta h_1) \; \times \Bigg(\frac{I_{n_1 + n_2 - n_3 - n_4 + 2l - 1}(\beta h) + I_{n_1 + n_2 - n_3 - n_4 + 2l + 1}(\beta h)}{2} \Bigg) \right]. \\
    \end{split}
\end{equation}

We can also compute the nematic magnetization $M_1 = P_1/Z$ with respect to the external field $h_1$ and its impure tensor is given as: 
\begin{equation}
\label{eq:nematic_mag}
    \begin{split}
        \mathcal{I}_{n_1 n_2 n_3 n_4}^{M_{1}} = &\;  \left[\sqrt{\prod_{k = 1}^{4} a_{n_k}(\beta,\Delta)} \; \times \sum_{l = -\infty}^{+\infty} I_{n_1 + n_2 - n_3 - n_4 + 2l}(\beta h)  \; \times \Bigg(\frac{I_{l+1}(\beta h_1) + I_{l-1}(\beta h_1)}{2} \Bigg) \right]. \\
    \end{split}
\end{equation}

Using $M$ and $M_1$, we can compute the magnetic susceptibility for a range of external fields and locate the BKT and half-BKT transitions. 

\section{\label{app:AppB}Numerical data table}

The following table shows the data for the critical temperature computed using a real-space tensor network method. 

\begin{table}[H] 
\renewcommand{\arraystretch}{1.25}
\setlength{\tabcolsep}{18pt}
\centering
\begin{tabular}{|c|c|c|c|} 
 \hline
 $\Delta$ & $T_{\rm{BKT}; h, h_1 \to 0}$ & $T_{\rm{BKT}, \rm{C_{v}}}$ & $T_{\rm{Ising},\rm{C_{v}}}$ \\ 
 [0.5ex] 
 \hline
0.04 & 0.880(1) & 1.00(1) & 0.08(1)\\
\hline      
0.08 & 0.843(5) & 0.95(1) & 0.17(1)\\
\hline      
0.12 & 0.803(5) & 0.90(1) & 0.26(1)\\
\hline      
0.16 & 0.773(5) & 0.86(1) & 0.34(1)\\
\hline      
0.20 & 0.755(7) & 0.82(2) & 0.42(2)\\
\hline      
0.24 & 0.725(7) & 0.78(2) & 0.50(2)\\
\hline      
0.28 & 0.678(3) & 0.74(2) & 0.58(2)\\
\hline      
0.30 & 0.690(10) & 0.72(1) & 0.62(1)\\
\hline      
0.32 & 0.685(8) & 0.705(5) & 0.64(2)\\
\hline      
0.34 & 0.694(2) & - & 0.685(5)\\
\hline      
0.36 & 0.720(1) & - & 0.715(5)\\
\hline      
0.38 & 0.735(6) & - & 0.735(5)\\
\hline      
0.40 & 0.756(3) & 0.74(1) & -\\
\hline      
0.45 & 0.785(1) & 0.78(1) & -\\
\hline      
0.50 & 0.798(8) & 0.815(3) & -\\
\hline      
0.55 & 0.833(1) & 0.84(1) & -\\
\hline      
0.60 & 0.849(1) & 0.86(1) & -\\
\hline      
0.65 & 0.853(5) & 0.88(1) & -\\
\hline      
0.72 & 0.886(1) & 0.91(1) & -\\
\hline      
0.80 & 0.890(4) & 0.95(1) & -\\
 [1ex] 
 \hline
\end{tabular}

\caption{The summary of the numerical results obtained in this paper. $T_{\rm{Cv}}$ is the critical temperature determined from the peak of specific heat with no external magnetic field, whereas $T_{h, h_{1} \to 0}$ is the critical temperature determined from the peak of magnetic susceptibility in the limit of vanishing external magnetic field, $h$ and $h_1$ respectively. We use the symmetry breaking field $h$ for $q = 1$ ($\Delta \in [0.38, 0.80]$) and $h_1$ for $q = 2$ ($\Delta \in [0.04, 0.36]$) in Eq.~\eqref{eq:appendixH} to compute the critical temperature.} For $T > 0.38$, there is a single transition of the BKT universality class. Until $\Delta = 0.32$, we can resolve the half-BKT and Ising line, but for $\Delta = 0.34, 0.36$ it is likely, based on our numerical results, that the two transition lines i.e., Ising and half-BKT have merged. For $\Delta \ge 0.40$, there is no ambiguity and transition corresponds to the BKT class. If all the transition lines meet, they do so at $\Delta = 0.36(2)$. Our results are slightly more consistent with the picture that first half-BKT and Ising lines meet around $\Delta \sim 0.34$, and then the Ising line continues to merge with the BKT line around $\Delta = 0.36(2)$.
\label{table1}
\end{table}

\section{\label{app:AppC}Collection of plots for range of \texorpdfstring{$\Delta$}{D}}
In this Appendix, we collect additional plots corresponding to the data in Table~\ref{table1}. 

\begin{minipage}{0.9\linewidth}
      \centering
      \begin{minipage}{0.45\linewidth}
          \begin{figure}[H]
              \includegraphics[width=\linewidth]{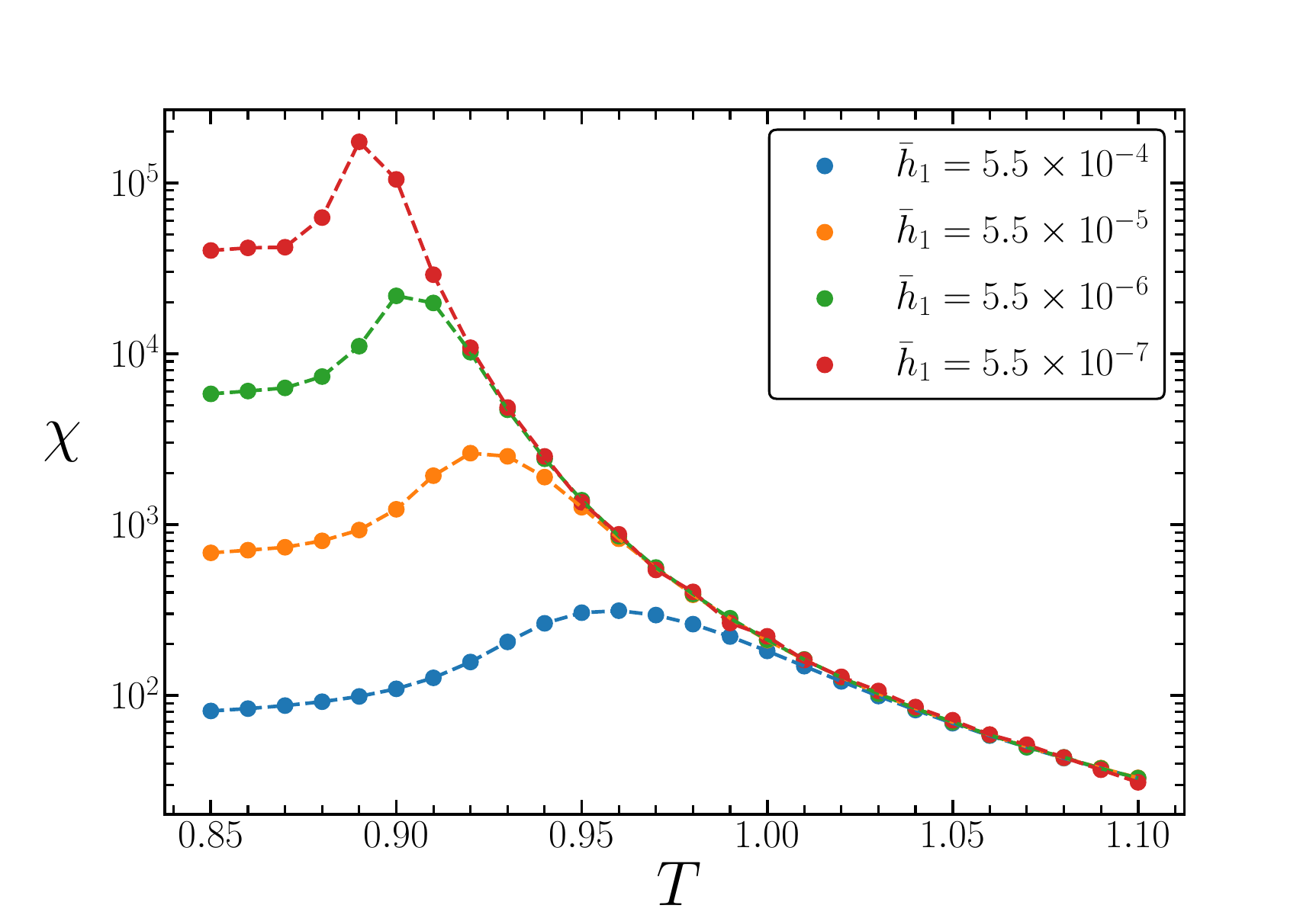}
              \caption{The variation of magnetic susceptibility, $\chi$, with temperature $T$ for $\Delta = 0.04$.}
              \label{fig:chi_T1}
          \end{figure}
      \end{minipage}
      \hspace{0.05\linewidth}
      \begin{minipage}{0.45\linewidth}
          \begin{figure}[H]
              \includegraphics[width=\linewidth]{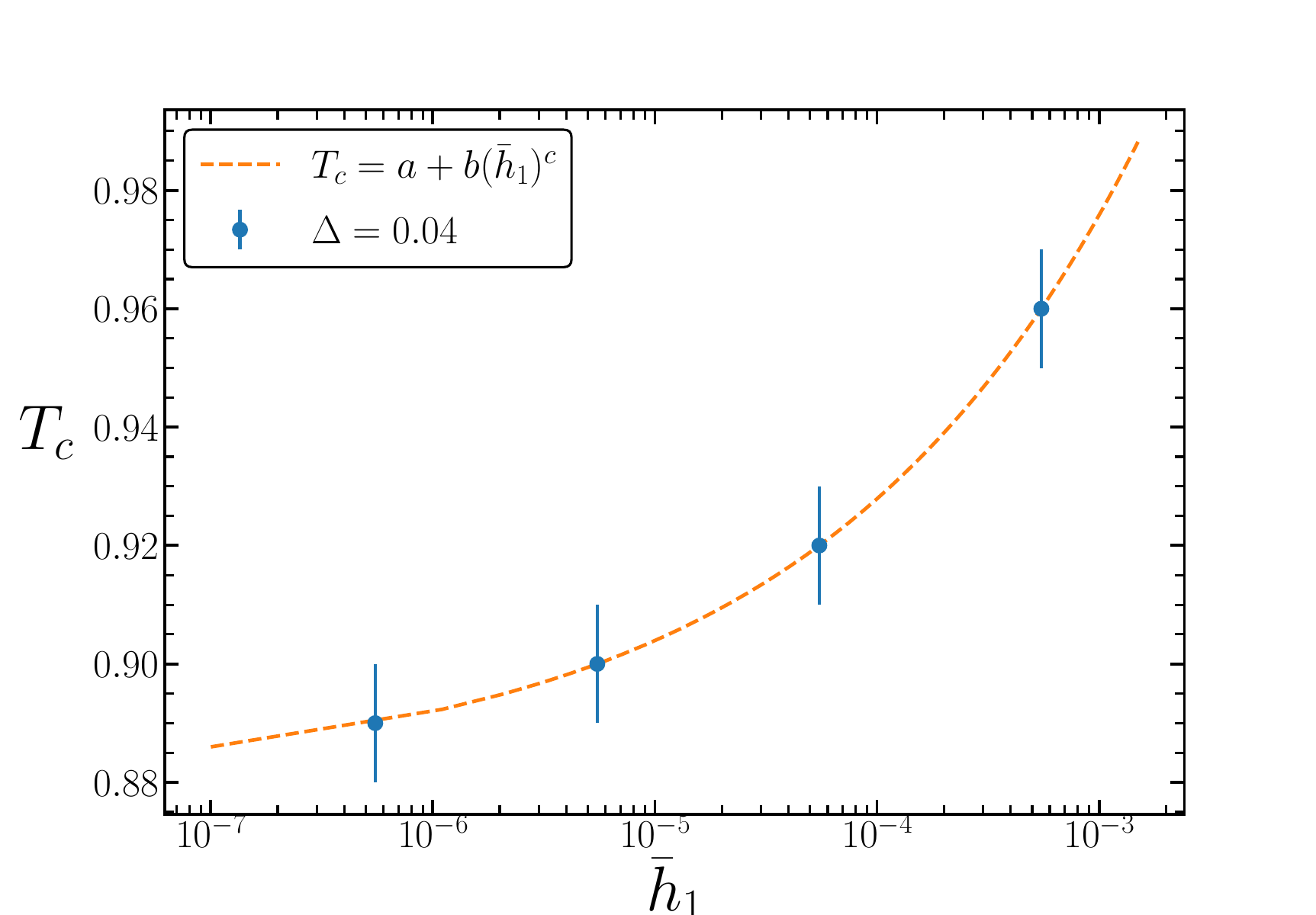}
              \caption{$T_c$ v/s $\overline{h}_{1}$ for $\Delta = 0.04$. $a = 0.880(1), b = 0.766(1), c = 0.301(1)$.}
              \label{fig:delta_T1}
          \end{figure}
      \end{minipage}
  \end{minipage}

  \begin{minipage}{0.9\linewidth}
      \centering
      \begin{minipage}{0.45\linewidth}
          \begin{figure}[H]
              \includegraphics[width=\linewidth]{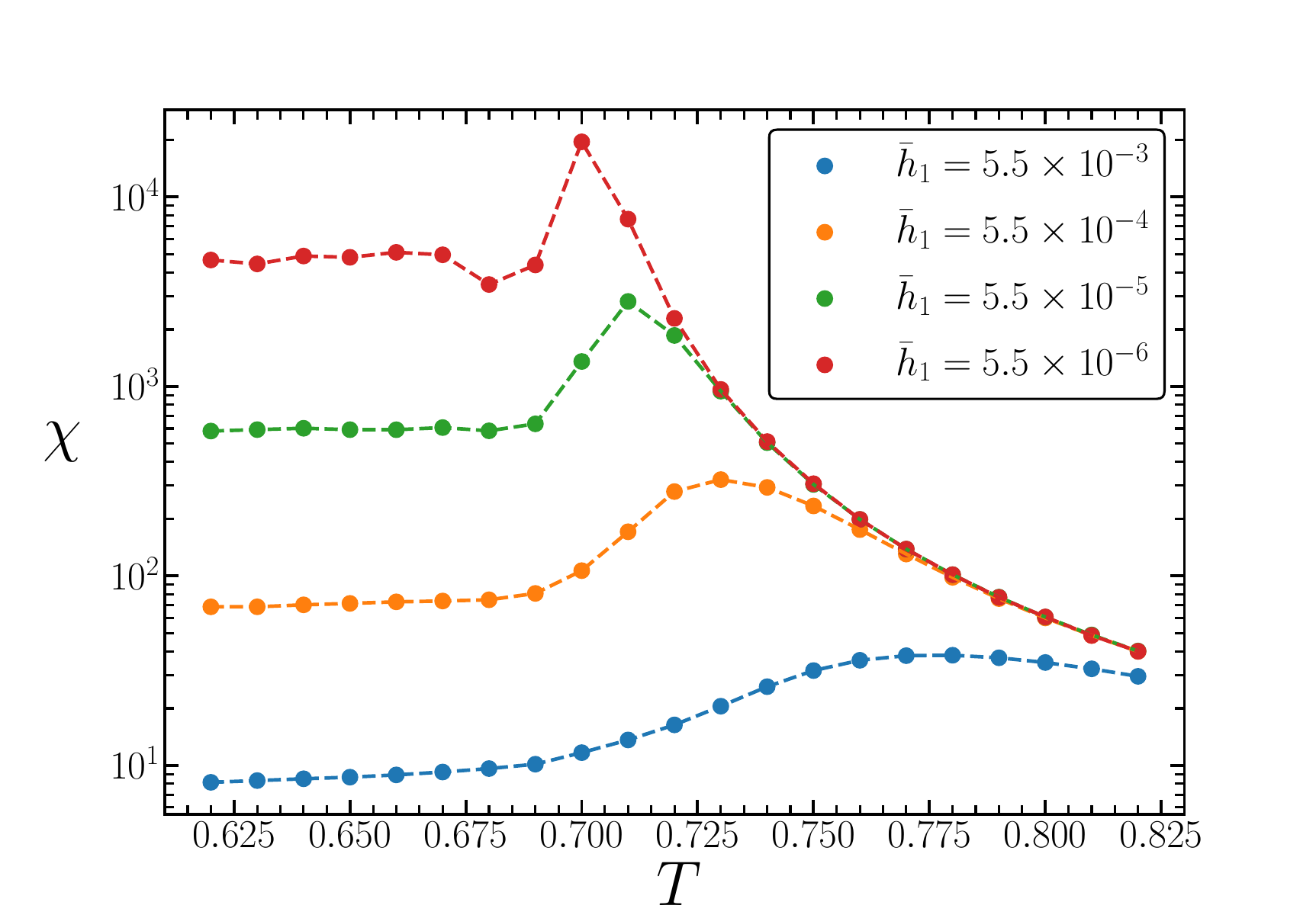}
              \caption{The variation of magnetic susceptibility, $\chi$, with temperature $T$ for $\Delta = 0.34$.}
              \label{fig:chi_T2}
          \end{figure}
      \end{minipage}
      \hspace{0.05\linewidth}
      \begin{minipage}{0.45\linewidth}
          \begin{figure}[H]
              \includegraphics[width=\linewidth]{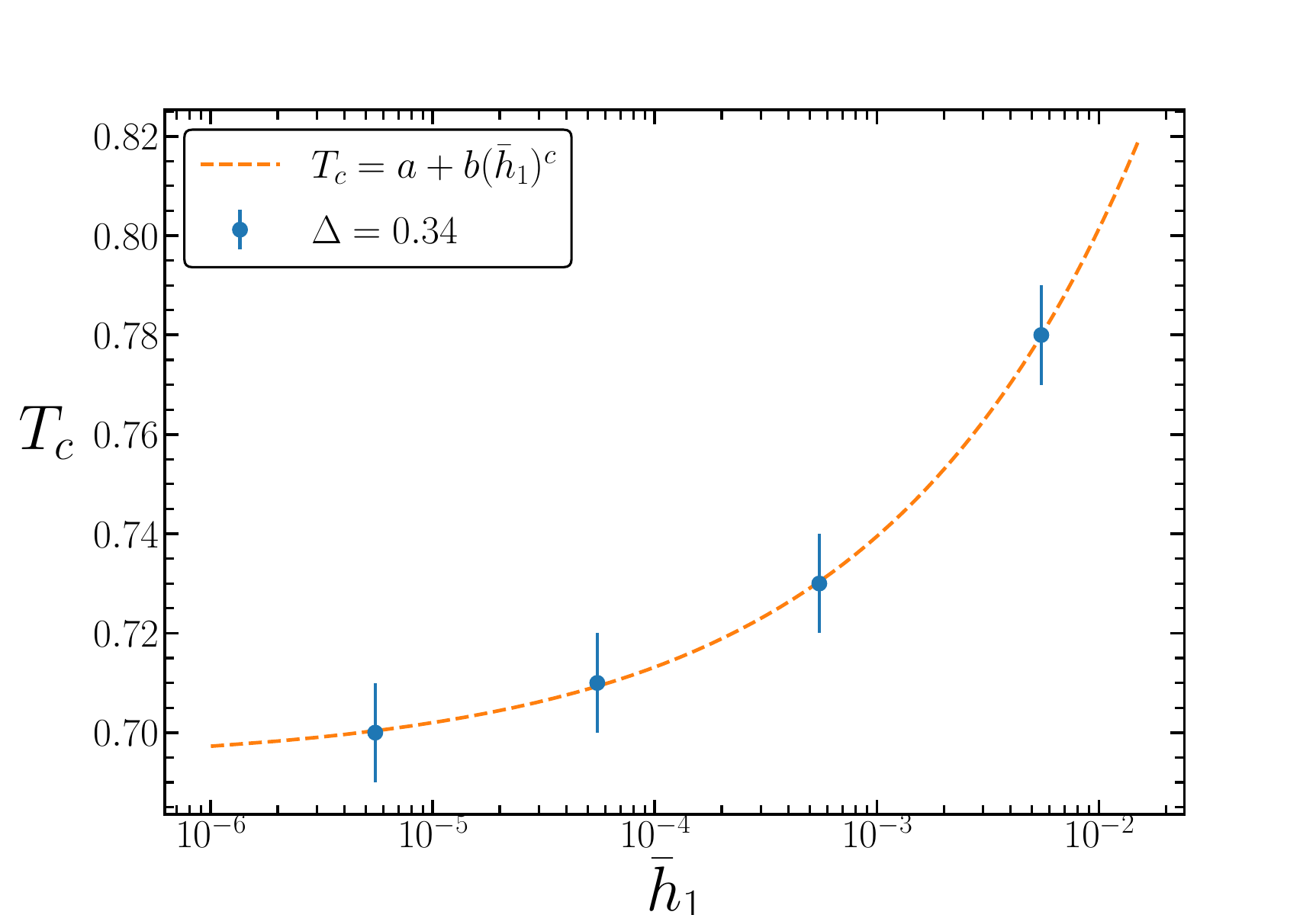}
              \caption{$T_c$ v/s $\overline{h}_{1}$ for $\Delta = 0.34$. $a = 0.694(2), b = 0.60(5), c = 0.37(2)$.}
              \label{fig:delta_T2}
          \end{figure}
      \end{minipage}
  \end{minipage}

  \begin{minipage}{0.9\linewidth}
      \centering
      \begin{minipage}{0.45\linewidth}
          \begin{figure}[H]
              \includegraphics[width=\linewidth]{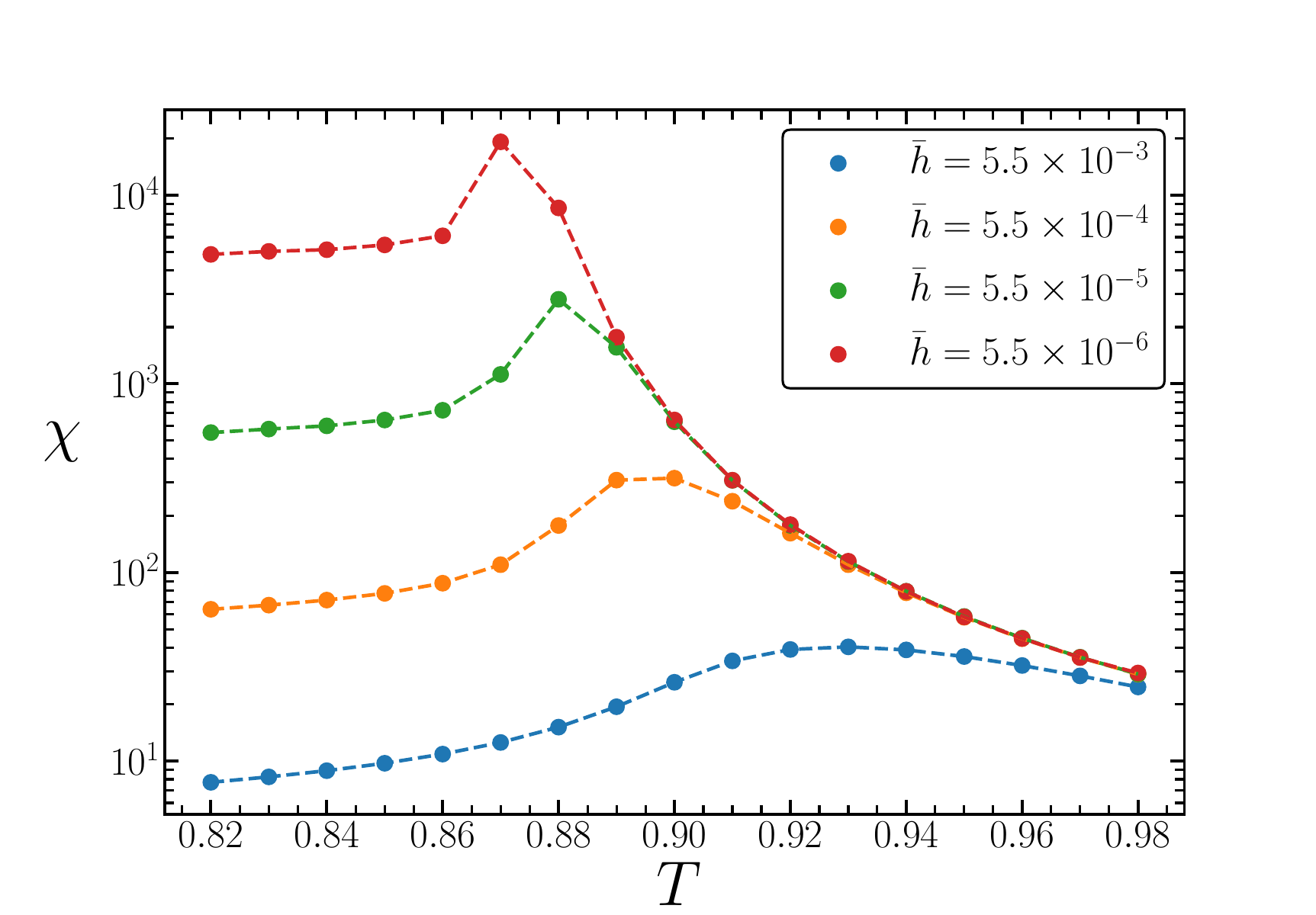}
              \caption{The variation of magnetic susceptibility, $\chi$, with temperature $T$ for $\Delta = 0.65$.}
              \label{fig:chi_T3}
          \end{figure}
      \end{minipage}
      \hspace{0.05\linewidth}
      \begin{minipage}{0.45\linewidth}
          \begin{figure}[H]
              \includegraphics[width=\linewidth]{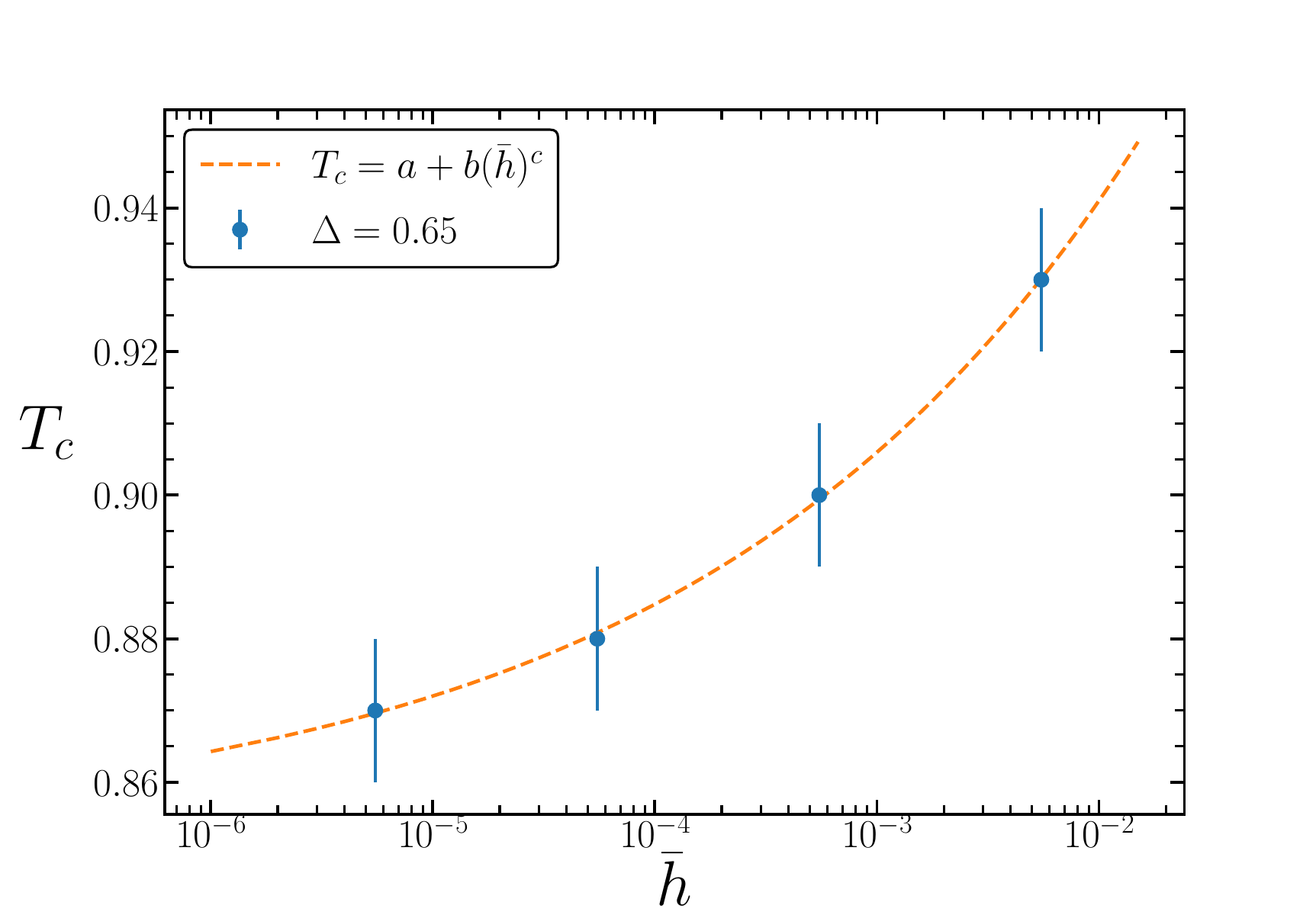}
              \caption{$T_c$ v/s $\overline{h}$ for $\Delta = 0.65$. $a = 0.853(5), b = 0.24(2), c = 0.22(3)$.}
              \label{fig:delta_T3}
          \end{figure}
      \end{minipage}
  \end{minipage}

\begin{minipage}{0.9\linewidth}
      \centering
      \begin{minipage}{0.45\linewidth}
          \begin{figure}[H]
              \includegraphics[width=\linewidth]{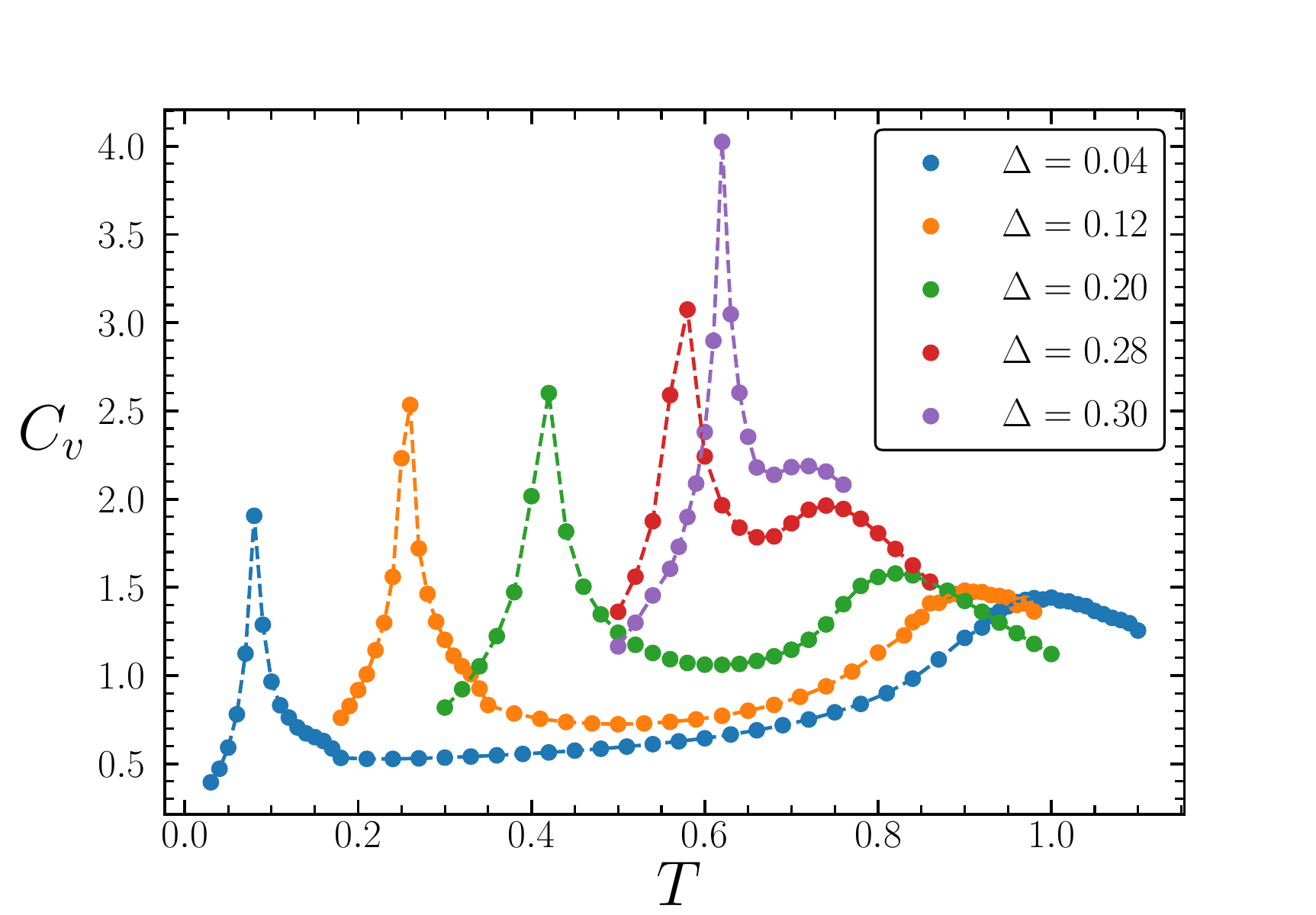}
              \caption{The specific heat, $C_{\rm{v}}$ with $T$ for $\Delta < 0.32$, $h, h_1 = 0$.}
              \label{fig:chi_T4}
          \end{figure}
      \end{minipage}
      \hspace{0.05\linewidth}
      \begin{minipage}{0.45\linewidth}
        \begin{figure}[H]
              \includegraphics[width=\linewidth]{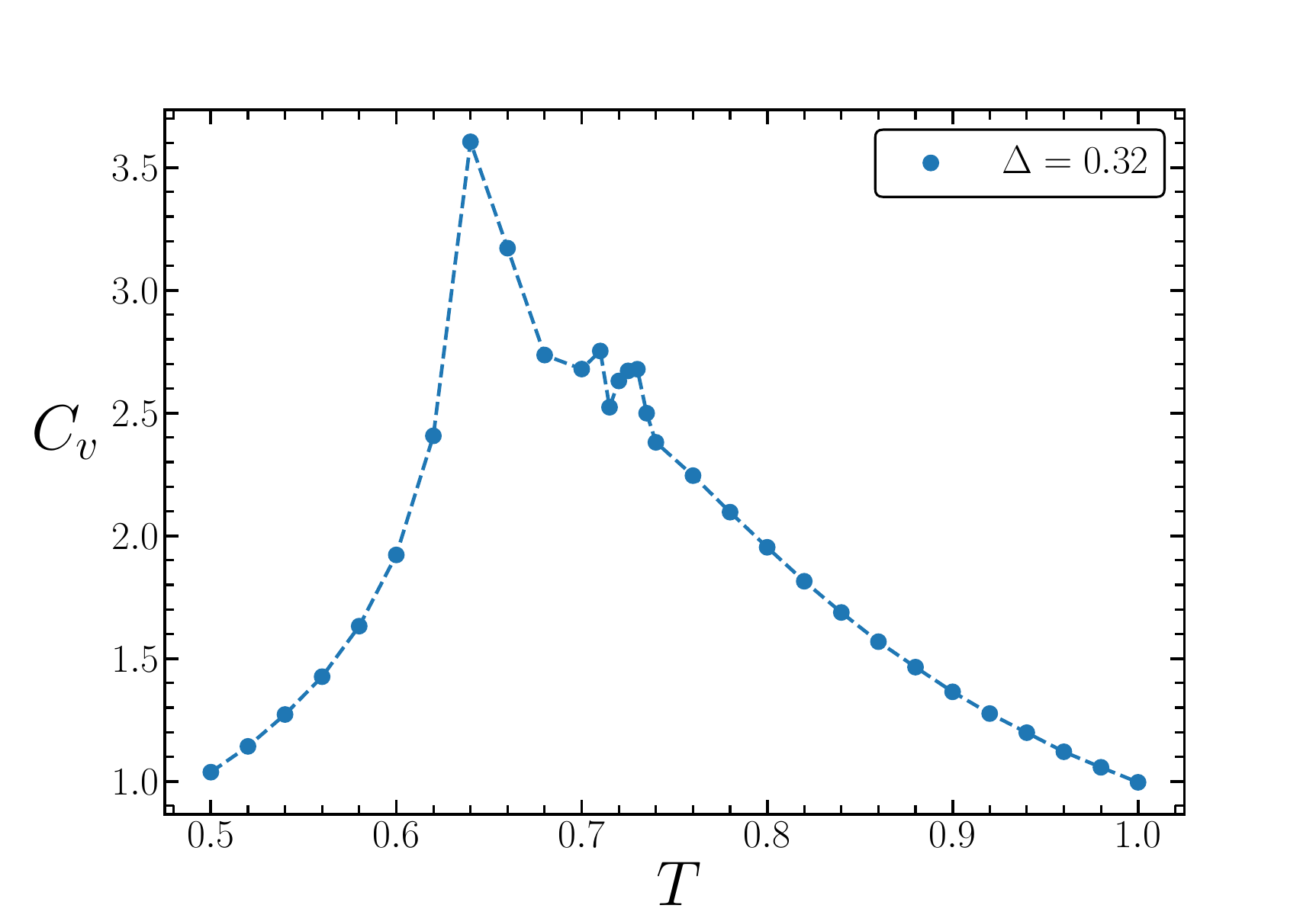}
              \caption{The specific heat, $C_{\rm{v}}$ with $T$ for $\Delta = 0.32$, $h, h_1 = 0$.}
              \label{fig:delta_T4}
          \end{figure}
      \end{minipage}
  \end{minipage}

\begin{minipage}{0.9\linewidth}
      \centering
      \begin{minipage}{0.45\linewidth}
          \begin{figure}[H]
              \includegraphics[width=\linewidth]{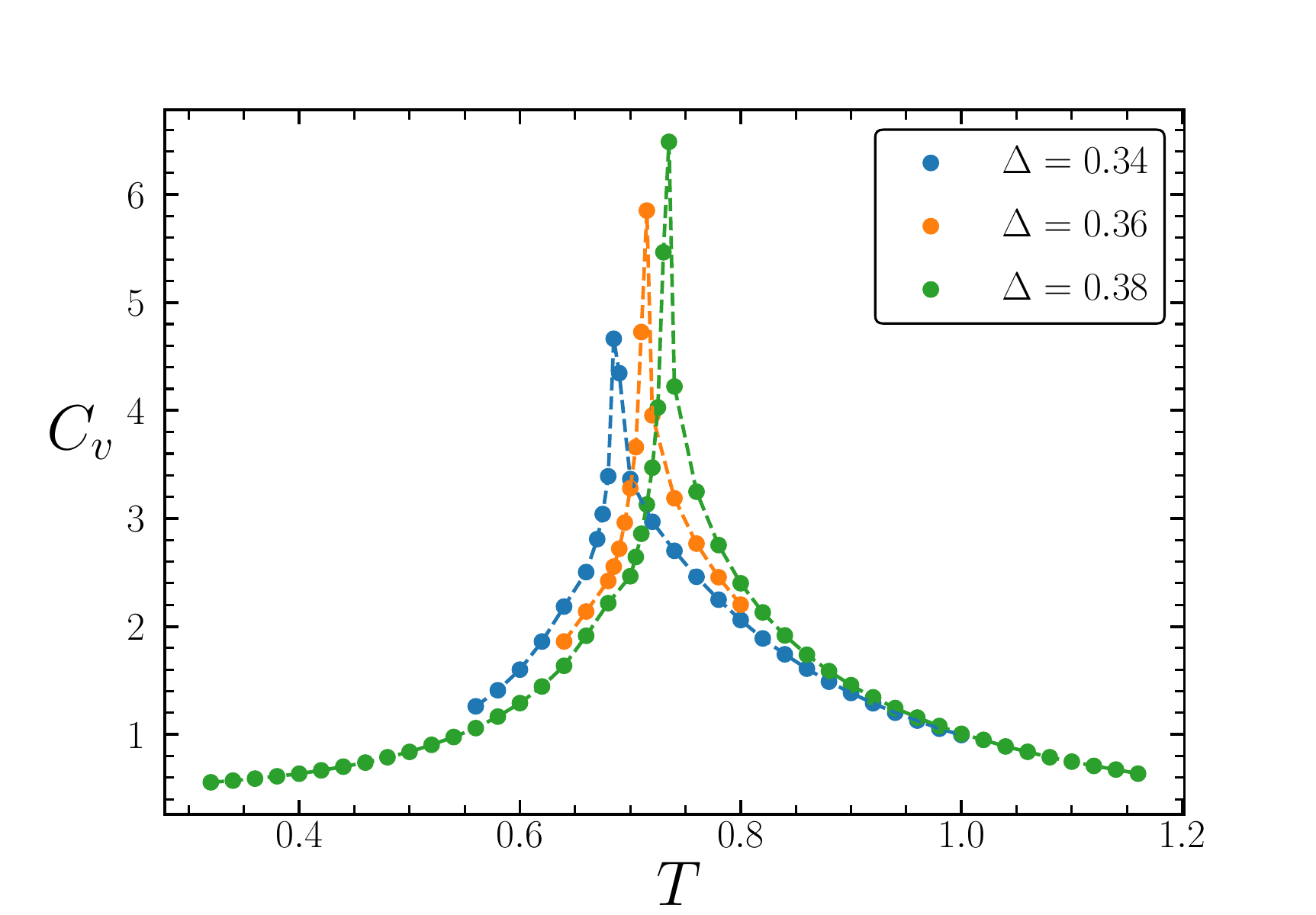}
              \caption{The specific heat, $C_{\rm{v}}$ with $T$ for $\Delta \in [0.34,0.4)$, $h, h_1 = 0$.}
              \label{fig:chi_T5}
          \end{figure}
      \end{minipage}
      \hspace{0.05\linewidth}
      \begin{minipage}{0.45\linewidth}
      \begin{figure}[H]
              \includegraphics[width=\linewidth]{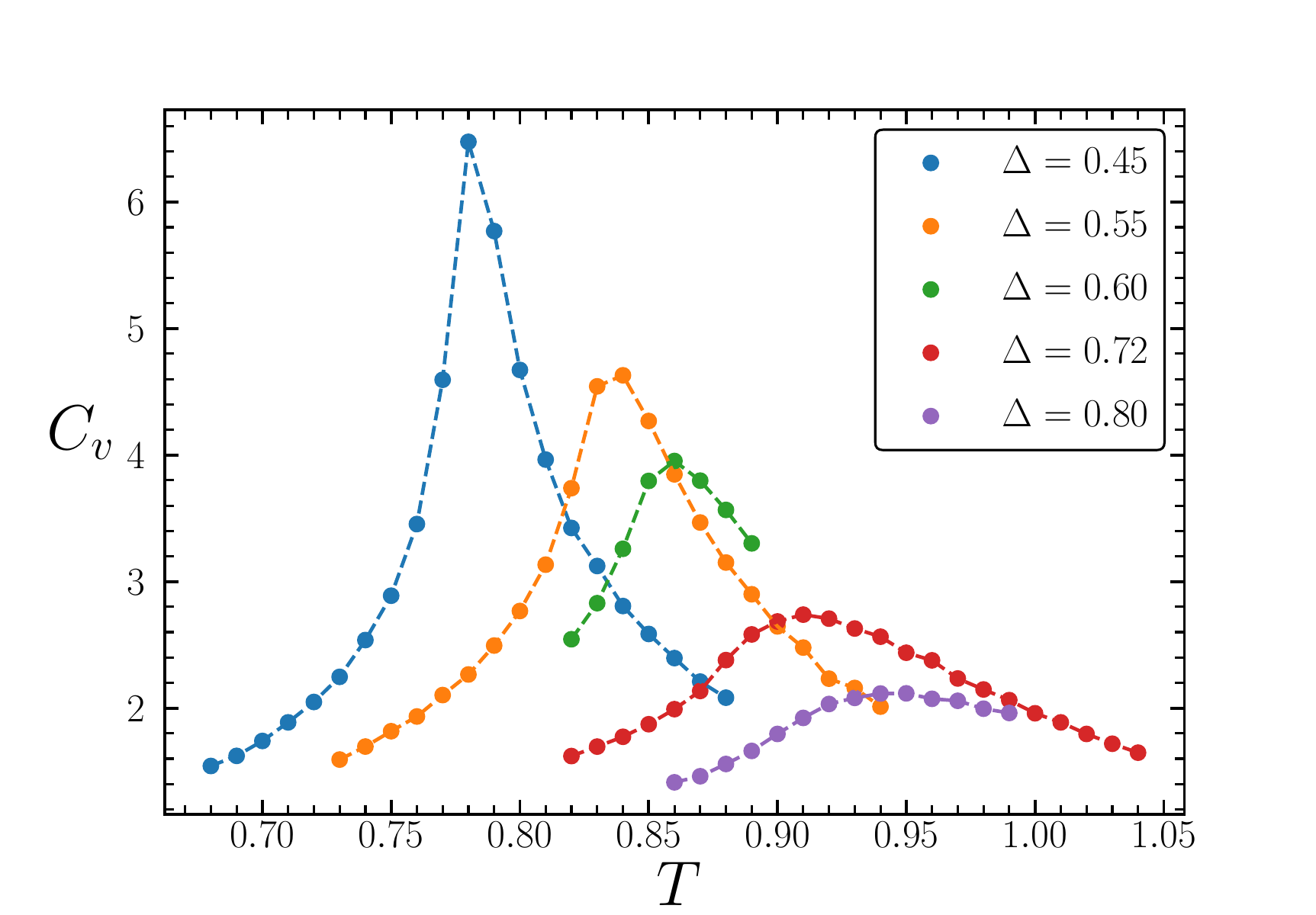}
              \caption{The specific heat, $C_{\rm{v}}$ with $T$ for $\Delta > 0.40$, $h, h_1 = 0$.}
              \label{fig:delta_T5}
          \end{figure}
      \end{minipage}
  \end{minipage}

\section{\label{app:AppD}Systematic Error Analysis for \texorpdfstring{$m$}{m} and \texorpdfstring{$D$}{D}}

The following table shows that free energy and magnetization do not show much variation as the range of allowed $m$ values increase at fixed volume, $\Delta, T \hspace{1mm} {\rm and} \hspace{1mm} D$.
\begin{table}[H] 
\renewcommand{\arraystretch}{1.25}
\setlength{\tabcolsep}{16pt}
\centering
\begin{tabular}{|c|c|c|} 
 \hline
 $m$ & $F$ & $M_1$ \\ 
 [0.5ex] 
 \hline
    [-20, 20] & -0.5436041  & 0.486666  \\ 
    \hline
    [-30, 30] & -0.5436043  & 0.486616  \\ 
    \hline
    [-40, 40] & -0.5436043  & 0.486624  \\ 
    \hline
    [-50, 50] & -0.5436043  & 0.486863  \\ 
    \hline
    [-60, 60] & -0.5436046  & 0.486586  \\ 
    \hline
    [-70, 70] & -0.5436066  & 0.486928  \\ 
    \hline
    [-80, 80] & -0.5436041  & 0.486907  \\ 
    [1ex]
    \hline
\end{tabular}

\caption{The free energy and magnetization computed using $h_1$ field for $\Delta = 0.34$, $T \approx T_c = 0.694$ and lattice volume of $2^{30} \times 2^{30}$ for different range of values for $m$ with $D = 91$.}
\label{table2}
\end{table}

\vspace{10mm}

The following plots show the scaling of free energy with bond dimension $D$ for different values of $\Delta$ and its corresponding critical temperature.

\begin{minipage}{0.9\linewidth}
      \centering
      \begin{minipage}{0.45\linewidth}
          \begin{figure}[H]
              \includegraphics[width=\linewidth]{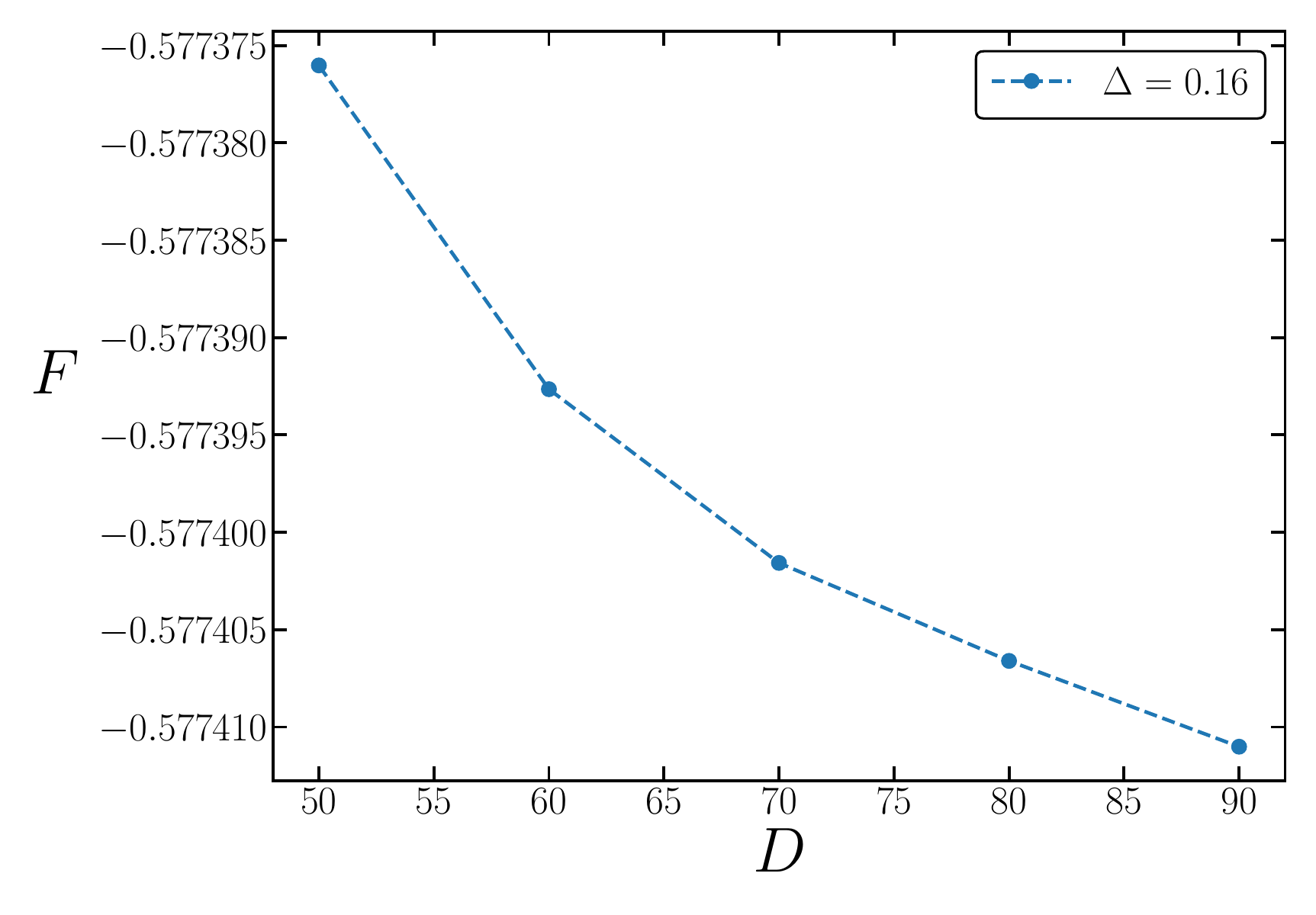}
              \caption{$D$-scaling for $\Delta = 0.16$, $m \in [-50,50]$, $T \approx T_c = 0.773$ and volume is $2^{30} \times 2^{30}$}
              \label{fig:scale1}
          \end{figure}
      \end{minipage}
      \hspace{0.05\linewidth}
      \begin{minipage}{0.45\linewidth}
      \end{minipage}
  \end{minipage}

\begin{minipage}{0.9\linewidth}
      \centering
      \begin{minipage}{0.45\linewidth}
          \begin{figure}[H]
              \includegraphics[width=\linewidth]{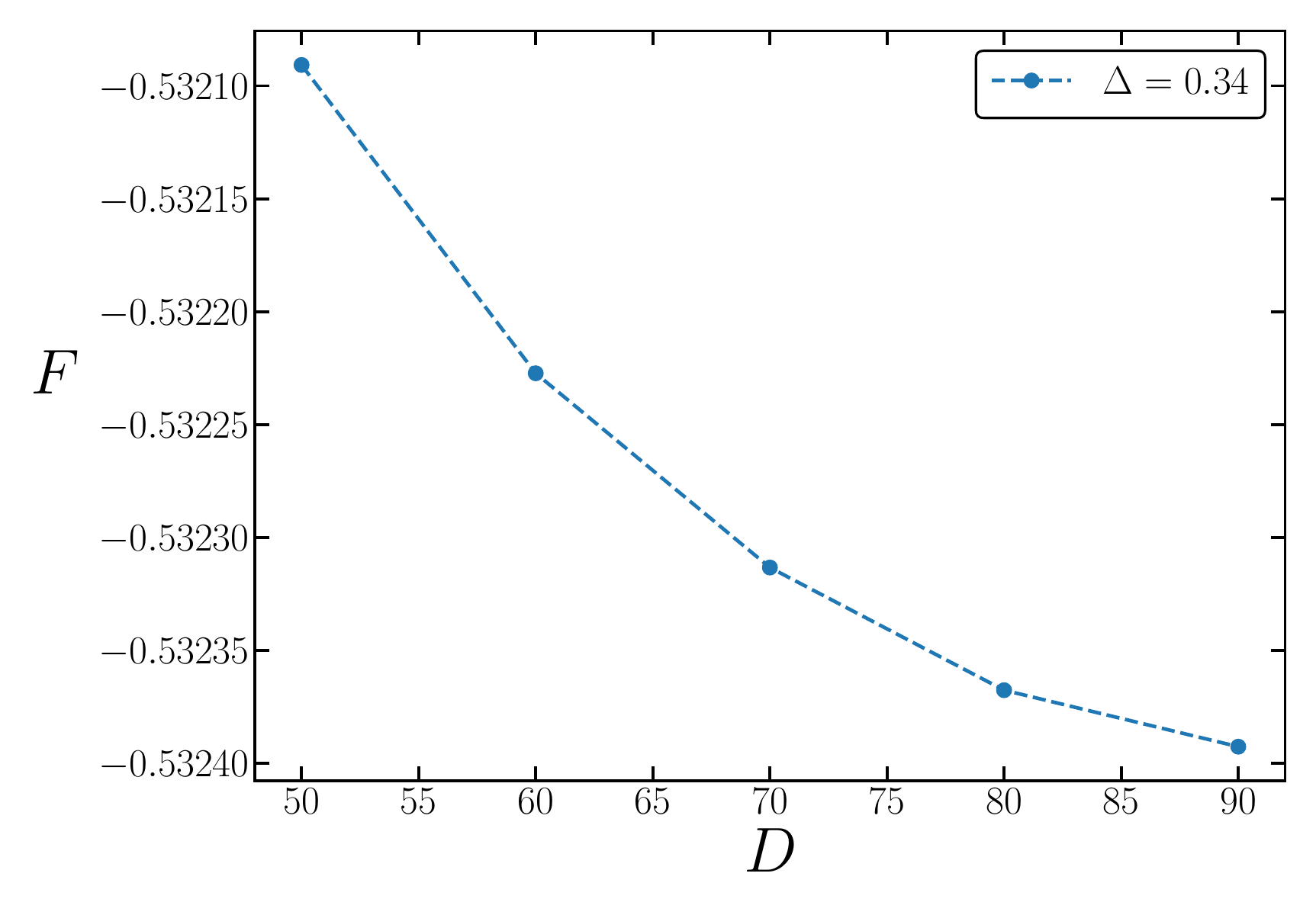}
              \caption{$D$-scaling for $\Delta = 0.34$, $m \in [-50,50]$, $T \approx T_c = 0.694$ and volume is $2^{30} \times 2^{30}$}
              \label{fig:scale2}
          \end{figure}
      \end{minipage}
      \hspace{0.05\linewidth}
      \begin{minipage}{0.45\linewidth}
      \begin{figure}[H]
              \includegraphics[width=\linewidth]{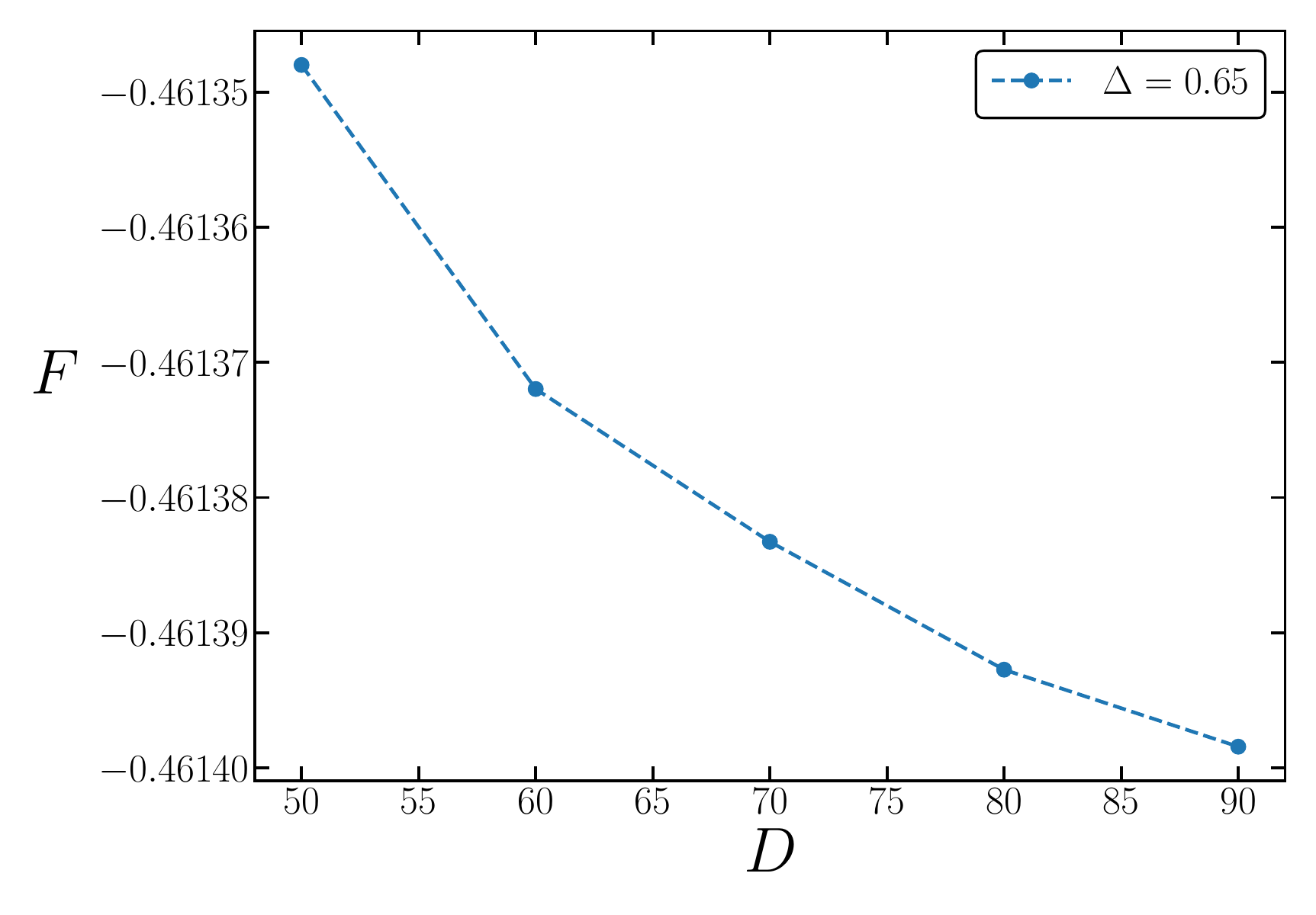}
              \caption{$D$-scaling for $\Delta = 0.65$, $m \in [-50,50]$, $T \approx T_c = 0.853$ and volume is $2^{30} \times 2^{30}$}
              \label{fig:scale3}
          \end{figure}
      \end{minipage}
  \end{minipage}

\vspace{10mm} 

\bibliographystyle{utphys}
\raggedright
\bibliography{main.bib}
\end{document}